# Local, Interactive, and Actionable: a Pandemic Behavioral Nudge

Alex Rich, MPH, MBA*, Cameron Yick, BS†, David Gotz, PhD‡

University of North Carolina at Chapel Hill, Datadog, University of North Carolina at Chapel Hill

**ABSTRACT**

The informational environment surrounding the Covid-19 pandemic has been widely recognized as fragmented, politicized, and complex [1]. This has resulted in polarized public views regarding the veracity of scientific communication, the severity of the threat posed by the virus, and the necessity of nonpharmaceutical interventions (NPIs) which can slow the spread of infections [2]. This paper describes CovidCommitment.org, an effort toward enhancing NPI adoption through the combination of a social behavioral commitment device and interactive map-based visualizations of localized infection data as tabulated via a 1-hour-drive-time isochrone. This paper describes the system design and presents a preliminary analysis of user behavior within the system.

**Index Terms**: [Human-centered computing]: Visualization

## 1 INTRODUCTION

More than 600,000 Americans have died of Covid-19 since the beginning of the pandemic in early 2020 [3]. A Lancet commission of researchers estimated that the US death count would have been 40% lower had it paralleled rates in other G7 nations [4]. Infection and death rates could have been reduced by effective public health policy and more timely adoption of nonpharmaceutical interventions (NPIs). NPIs are often the only infection prevention strategies available prior to vaccine development. These include limitations on public gathering, closures of businesses and schools, and enaction of stay-at-home orders and mask mandates [2]. Public messaging during the pandemic was chaotic and largely split along political party lines [5]. Some state and national political leaders consistently downplayed the severity of the pandemic and the potentially deadly consequences of becoming infected [5]. The informational environment surrounding Covid-19 became known by many as an "infodemic," with a high volume of confusing and sometimes conflicting information flowing to citizens through a variety of channels [1]. This contributed to popular rejection of NPIs that could have blunted the spread of Covid-19 in regions throughout the US [6].

Research conducted by a team assembled under the George W. Bush administration demonstrated that the early implementation of multiple NPIs during the 1918 Spanish Flu pandemic had the effect of reducing peak death rates by 50% [2]. For example, St. Louis enacted NPIs 14 days faster than Philadelphia relative to the first reported infection in each city, and the early response correlated with a remarkable reduction in peak death rate (Figure 1).

*alex.rich@unc.edu
†cameron.yick@gmail.com
‡gotz@unc.edu

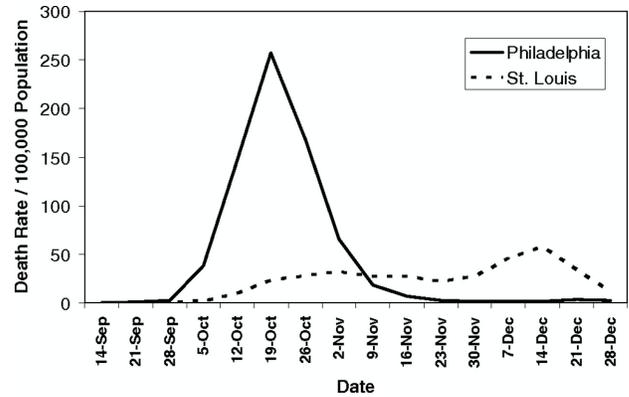

Figure 1: "Curve Flattening" effect of early nonpharmaceutical interventions in St. Louis during the 1918 pandemic [2].

Unfortunately, enacting NPIs involves significant sacrifice on the part of the public and NPIs are most effective early in an outbreak when infection rates appear low and the threat to the populace is least defined. This can drive strong resistance to NPI implementation at the moments in which implementation is most effective. This resistance was typified by the public comments of the Lieutenant Governor of Texas in March 2020, when he advocated the sacrifice of lives among the elderly in order to avoid the economic implications associated with aggressively initiating NPIs [7]. At the time, the majority of the publicly available Covid-19 infection count information was aggregated to the national or state level, making it difficult for members of the general public to develop a sense of whether infections were occurring near their homes. Many communities thus faced a situation in which the threat of Covid-19 infection was abstract and lacked saliency, while the economic cost of NPI implementation was personal and tangible.

Motivated by these challenges, we designed CovidCommetment.org, a social behavioral commitment device that includes an interactive map-based visualization using isochrones to personalize Covid-19 data to a user's locale. This project was undertaken to enhance the uptake of NPIs among the American population through enhanced communication of information on infections, deaths, and NPIs in users' local areas. Texas residents were the initial target audience prior to expanding to the rest of the US. The design was loosely patterned after aviation instrument panels, aiming to provide a simple, standardized display of only the most important local data. It was designed to allow lay users to answer the questions "What is the virus doing near me?" and "What can I do to stay safe?".

## 2 RELATED WORK

Zhang et al. completed an extensive review of the many forms of data visualization that emerged around the pandemic [8]. Their work outlines numerous categories of pandemic visualizations and documents the emergence of national-level bubble maps and choropleths that were prevalent as we developed and prototyped CovidCommitment.org. Within their framework, our intervention would be classified as a nongovernmental volunteer organization

("who") utilizing the Johns Hopkins Center for Systems Science and Engineering infection and death count repository [3] ("what data") to communicate local infection and death counts ("what message") in interactive line plots, geographic visualizations, and a text-and-icon commitment device ("in what form") to lay users without epidemiological or statistical training ("to whom") with the intent to encourage adoption of CDC recommended NPIs plus an additional item regarding maintaining social connections via digital means ("with what effect").

Isochrone-based data visualization for public health applications has not received much attention in the literature. Zeng et al. presented data visualizations evaluating public transportation utilizing isochrones in 2014 [9]. The work presented here extends this concept through the use of isochrones to calculate and visualize local infection and death data rather than transportation accessibility.

Research on visual analytics for public health applications often focuses on expert users seeking to understand outbreak dynamics and risk factors. Preim and Lowonn provide an excellent review of the analytical approaches utilized by experts [10]. Their work illustrates the level of complexity that arises in tools developed by public health professionals for use by fellow public health professionals. Tools designed for lay users need to be less complex in order to encourage engagement and reduce confusion.

## 3 THEORY

The work being presented rests on theory drawn from multiple disciplines. Crisis risk communication informed the informational content and framing. The commitment device at the center of the work was drawn from behavioral science. Finally, the design of the web application was grounded in principles from the field of data visualization and interaction design.

### 3.1 Crisis Risk Communication

Crisis risk communication is a multifaceted discipline that draws from multiple fields [11]. Research in the field highlights gaps between the ways in which scientists evaluate information and the ways in which members of the public form perceptions and respond to the information presented to them [12]. It also provides a number of frameworks for the design and evaluation of information sharing strategies in the context of significant public health threats.

In managing the intersection between epidemiological expertise and the perception of the general public, it is helpful to view the problem through the lens of "hazard" and "outrage," where hazard represents the mathematical rate of dangerous events and outrage represents the emotional response among the public that influences the way in which risks are perceived [13]. There are many factors at play in the emotional response to a threat like a pandemic. The factors of particular interest in the spring and summer of 2020 were related to the perceived trustworthiness of government agencies, the sense of personal control vs. government-mandated restrictions, and the saliency issues associated with the most frightening aspects of the disease being hidden from public view in hospitals and diffused in time and space.

Guidance from the US Centers for Disease Control and Prevention (CDC) provides a lens through which to design and evaluate public health communication efforts during a pandemic. The CDC approach to dissemination of information to the public during a pandemic is outlined in its Crisis and Emergency Risk Communication (CERC) manual [14]. It centers on six principles: (1) Be First, (2) Be Right, (3) Be Credible, (4) Express Empathy, (5) Promote Action, & (6) Show Respect. These principles helped guide the design for CovidCommitment.org.

### 3.2 Behavioral Science

The field of behavioral science provides a means of understanding the biases and heuristics associated with decision making and a set of tools for nudging individuals toward desirable behavior. NPIs represented significant changes in the daily lives of Americans and faced resistance among broad sections of the population. NPI implementation thus required individuals to commit to making a different set of choices from what they were accustomed to multiple times each day. The MINDSPACE pneumonic developed by Dolan et al. provides a helpful framework for considering the behavioral science associated with this kind of problem [15]. Of particular interest for this project were elements related to messenger, salience, and commitment devices. The application was intended to be as politically neutral as possible while enhancing the salience of the threat information being presented through localization. This impacted data source selection and the final formats of information displayed. The commitment device became a central element of the final design. Commitment devices are any means of formalizing an intent to adhere to desired behavior in the future. They can involve penalties for non-adherence or social pressure, but the simple act of writing down a commitment has been shown to have beneficial effects [16].

### 3.3 Data Visualization

Information visualization can enhance the speed of perception and clarity of insights in the data being presented. Visualization offers the ability to do some of the cognitive work for the user, easing the path toward absorption of the desired insight [17]. In developing a means of rapidly communicating explanatory information about the local Covid-19 infection hazard, data visualization can be utilized to improve user engagement and reduce the rate at which users abandon the site without interacting.

John Snow made his case for the waterborne transmission of cholera by mapping clusters of cholera deaths around a single water pump during an outbreak in London in 1854 (Figure 2) [18]. Though he had long suspected waterborne transmission and his published work on the matter was far more oriented toward text and tabular data, he is remembered for his data visualization: the map.

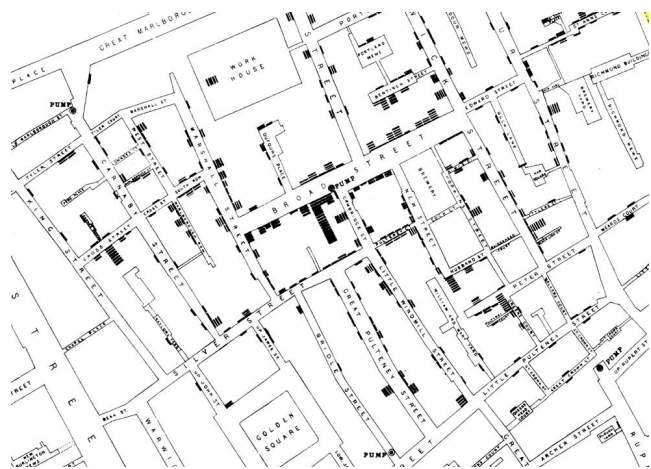

Figure 2: John Snow's 1854 cholera outbreak map showing proximity of deaths (black rectangles) to the Broad Street water pump [19]

Geographic data visualization remains a key element of public health communication and a valuable means of personalizing information presentation. Literature on suggested interactivity provides a framework for discussing design elements that encourage discovery of interactive elements on a web application [20]. Discoverability and feedforward effects were important elements of the final design.

## 4 DATA AND METHODS

The team set out to design a tool which would increase the saliency of the Covid-19 threat to users and drive implementation of NPIs by communities outside the initial outbreak areas. The initial prototype was built in a Jupyter notebook and later converted into a web-based React application. This effort involved sourcing data on Covid-19 impacts, developing a means of displaying elements of that data as local-to-the-user information, and creating a call to action encouraging NPI uptake.

The "Be First" component of CDC CERC guidelines demands consistent updates to information being presented. There were two sources of daily updated infection and death count data available during application development in March of 2020. Both the New York Times and the Johns Hopkins Center for Systems Science and Engineering provided algorithmically accessible repositories aggregating local health department data at the county level [3], [21]. Either data source would enable the team to comply with CDC's "Be Right" guidance, but the "Be Credible" item required evaluation of potential user perspective regarding media sources. The political climate of the time drove the selection of Johns Hopkins data, which we hoped would be perceived as a more neutral source by users from both sides of the American political spectrum.

Collecting localized Covid-19 data required several steps. First, the user-provided location needed to be geocoded. This was accomplished via the GoogleMaps application programming interface (API) [22]. Next, the isochrone for a 1-hour drive time around the geocoded point was obtained via the Heidelberg Institute for Geoinformation Technology's openrouteservice API [23]. A script was written using the Python GeoPandas library to identify the counties intersected by the isochrone [24]. Infection and death data for the intersected counties were then joined with the time series data posted in the Johns Hopkins CSSE repository.

With the data collected, the isochrone and intersected counties were visualized using the Python Folium library for producing interactive leaflet.js maps [25], [26]. The resulting map provides the user with a visual reference for the geographic region from which data were drawn (Figure 3). It also allows access to individual cumulative county totals via tooltips accessible by mouseover. The maps auto-center and scale to the isochrone, but allow zoom interaction on both desktop and mobile.

Trends in infections and deaths within the intersected counties is provided via rolling 7-day raw counts and visualized as an interactive line plot (Figure 3) made with the Semiotic framework [27]. The team evaluated numerous formats for displaying infection and death data including rates, age-adjusted rates, measures of socioeconomic status, demographic information, and raw infection and death counts. Feedback from scientists, physicians, and public health professionals demonstrated strong preference for adjusted rate statistics. Feedback from users among the general population demonstrated that many experienced stronger reactions to raw counts (e.g. "500 infected people within an hour's drive of me in the last 7 days") than were generated with rate data (e.g. "3 people per 100,000 within an hour's drive of me were infected this week"). Because the project aimed to be understandable to a general audience without assuming a background in statistical reasoning, the team chose to display count data instead of rate data.

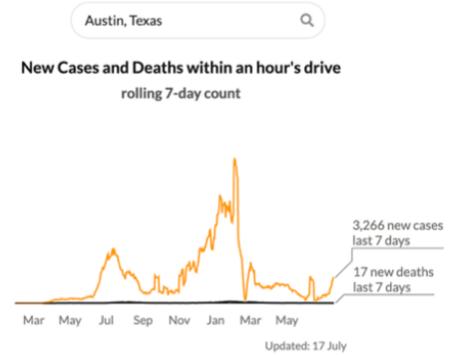

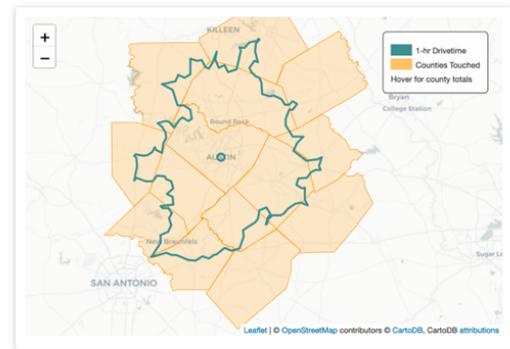

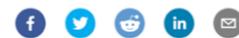

Figure 3: key elements of the CovidCommitment.org user interface

In keeping with CDC CERC guidelines, the next step of development focused on encouraging users to take action while expressing both respect and empathy. To the team, this meant communicating CDC guidelines for physical distancing and mask wearing in plain language while also recognizing the potential emotional challenges associated with loss of social contact. The original prototype simply listed these items below the visualizations. Consultation with behavioral scientists subsequently drove us to present the recommendations as an interactive commitment device based on expert feedback. This component enabled users to click checkboxes committing to each of five actions and provides the option to share their commitment with their social networks (Figure 3).

Bounce rate, a measurement of the proportion of users who abandon a site without interacting, was a major concern during the design process. The site incorporated multiple software interface design patterns that link to the field of interaction design in order to maximize engagement. Animation is used to guide and retain user attention when viewing the initial landing page, provide a sense of progress while awaiting results, to invite engagement with the commitment device. The landing page utilizes a cut off map image with hover-over tooltips alongside location input animation to provide a feed-forward suggestion of the interactive capability afforded. Each button or icon in the site changes color with hover-over to suggest interactive capabilities.

The initial landing page includes a prominent blinking cursor in the location input box and a rolling number display drawing the eye to the number of "Covid Commitments" made so far. Checkboxes next to commitment device items subtly pulse in size suggesting the potential for interactivity. A progress bar provides reassurance as the site collects data and prepares visualizations after location input. After viewing personalized data visualizations, unexpected confetti effects accompany each clicking interaction with a commitment device checkbox, driving the user to continue toward completing the commitment device with the large orange button at the bottom of the section. Clicking the final commitment button results in an alert box inviting the user to share their commitment on social media via built in functions. Finally, completion of the commitment sequence results in a change to the total count of Covid Commitments made, allowing the user to immediately see their impact on the site.

## RESULTS

The application was developed in March 2020 as a public service rather than a research project. The team's focus was on enhancing user engagement and reducing load times rather than on formal methods of evaluation and documentation of the site's performance as a behavioral nudge. The site has remained active for over a year, but received the most attention early in the pandemic. Results presented here cover the time period during which the site saw the most activity, April 1st 2020 through June 30th 2020. Without an advertising or marketing budget, CovidCommitment.org was viewed by 8,893 unique users in the US and 1,242 unique users in 83 other countries during that time. There were two peaks in user counts. First, there were 855 users on April 10th 2020 when the site was shared on Twitter by a user with a large following. Next, there were 907 users on May 19th 2020 when the site was recommended on the Today Show's website. Among the US user population, 2,457 returned to the site multiple times. Anecdotally, the team did receive feedback from when daily updates were delayed, suggesting that at least some users were persistently monitoring the pandemic via the site.

Users made 828 Covid Commitments during the high traffic time window. While there was no system in place to prevent a user from making multiple Covid Commitments, it is reasonable to assume that such behavior was rare. Assuming unique commitments, 8.2% of users in the time window completed the commitment device. Daily commitments peaked during April 8th 2020 at 74. Users shared their commitment on Facebook 214 times and Twitter 77 times.

## 5 DISCUSSION

The work discussed in this paper was limited in its evaluation methods by the challenges associated with coordination during the early phase of the Covid-19 pandemic. The team made efforts to expand the project to international users, but was unable to establish consistent sub-national data for the majority of non-US countries. A number of adjustments had to be written into the code to account for differences in reporting within the US including the geographic breakdown of data for New York City.

Further research on the use of this form of isochrone-enabled localized information presentation is called for. This work was based on the theory that resistance to NPI adoption was at least partially based lack of knowledge of infections and deaths in a person's local area. Resource limitations and timing did not allow for rigorous testing of the hypothesis that detailed local outbreak information could change attitudes toward NPIs. Future work should evaluate change in user attitudes before and after use of a tool similar to the one described. Larger studies should be designed to measure NPI adherence before and after using a tool like the one outline while gathering enough user demographic data to control for potential confounding effects.

## 6 CONCLUSION

The availability of free or low-cost APIs for creating isochrones opens new doors in public health research and communication. The volume of traffic despite the absence of advertising and high rate of engagement with the commitment device in CovidCommitment.org demonstrated demand for personalized local information during a public health emergency. Presentation of information user-centered time-to-travel frames of reference rather than arbitrary geographic boundaries may have significant impact on human behavior. Further research should evaluate this effect under more controlled conditions.

The project was unable to evaluate real world adoption of NPIs. However, it was able to demonstrate the technical feasibility of localizing data presentation to users and provide some sense of the popularity of such a localized display in a moment of crisis. We encourage future practitioners to consider applying some of the techniques that we applied in information displays (localized isochrone maps and behavioral commitment devices) as a potential means of increasing interest in and adherence to NPI guidelines during future communicable disease outbreaks.

## 7 ACKNOWLEDGMENTS

The authors would like to thank Dr. Dan Ariely of the Duke Center for Advanced Hindsight for his advice regarding behavioral science and commitment devices. We are also grateful to Dr. Sten Vermund, Dean of the Yale School of Public Health, whose advice on the inclusion of trending data and public support for the project greatly improved the interface and drove traffic to the site.